# Retraction of dissolution front in natural porous media

Yi Yang,[1*] Stefan Bruns,[1] Melania Rogowska,[1] Sepideh S. Hakim,[1] Jörg U. Hammel,[2] Susan L. S. Stipp[1] and Henning O. Sørensen[1]

[1] Nano-Science Center, Department of Chemistry, University of Copenhagen, Universitetsparken 5, DK-2100 Copenhagen, Denmark

[2] Helmholtz-Zentrum Geesthacht, Max-Planck-Straße 1, 51502 Geesthacht, Germany

* Corresponding author: yiyang@nano.ku.dk

*The dissolution of porous materials in a flow field shapes the morphologies of many geologic landscapes. Identifying the dissolution front, the interface between the reactive and the unreactive regions in a dissolving medium, is a prerequisite for studying dissolution kinetics. Despite its fundamental importance, the dynamics of a dissolution front in an evolving natural microstructure has never been reported. Here we show an unexpected spontaneous migration of the dissolution front against the pressure gradient of a flow field. This retraction stems from the infiltration instability induced surface generation, which can lead to a reactive surface dramatically greater than the ex situ geometric surface. The results are supported by a very good agreement between observations made with real time X-ray imaging and simulations based on static images of a rock determined by nanoCT. They both show that the in situ specific surface area of natural porous media is dependent on the flow field and reflects a balancing between surface generation and destruction. The reported dynamics challenge many long-held understanding of water-rock interactions and shed light on reconciling the discrepancies between field and laboratory measurements of reaction kinetics.*

In a reacting porous medium, a reaction front is an isosurface on which the chemical energy of a reaction reduces to zero in the velocity direction. This isosurface separates the reactive surface from the remaining geometric surface and is the ideal boundary of the region of interest (ROI) in the kinetics measurements of water-rock interactions. The inherent heterogeneities of natural porous materials that are not contained by this isosurface do not contribute to the chemical reaction. Therefore, the ability to track an evolving reaction front can greatly simplify the analysis of many geochemical processes by allowing us to identify unambiguously the temporal ROI. The dissolution front is the reaction front of a solid dissolution reaction. In a dissolving medium with a pressure driven flow field, there is a positive feedback between the mineral dissolution rate and the local permeability.[1, 2, 3, 4] This feedback leads to a morphological instability of the migrating dissolution front, referred to as reactive infiltration instability (RII).[5, 6] RII is considered a fundamental principle behind many geological self-organisation phenomena.[6, 7] However, the dynamics of a dissolution front in the presence of the RII in natural microstructures remain elusive,[8, 9, 10, 11, 12] presumably because of the technical challenge of observing it experimentally [13, 14, 15, 16] and effectively incorporating the inherent heterogeneities of porous media into numerical simulations.[5, 17, 18, 19, 20, 21, 22, 23]

**Surface generation by infiltration instability**

Given a flow field, the position of the dissolution front along any streamline can be determined using a plug flow approximation:[24]

$$\int_{pH_0}^{pH_{eq}} \frac{d[\text{H}^+]}{r_{diss}} = \int_{\mathbf{s}} \frac{SSA}{\mathbf{v}} ds, \qquad (1)$$

where [H$^+$] represents the hydronium cation concentration (mol·m$^{-3}$), pH$_0$ and pH$_{eq}$, the initial and equilibrium compositions of the rock-dissolving fluid (pH used as a master variable), $r_{diss}$ represents the mineral dissolution rate (mol·m$^{-2}$·s$^{-1}$) and typically depends on pH and the mineral saturation index (SI); SSA is the specific surface area (m$^2$/m$^3$), **v** the velocity and **s** the streamline as the integral path. Equation 1 relates the chemical potential of the fluid (left hand side, LHS) to the physical properties of the medium (right hand side, RHS). The LHS is constant given a specific reaction and initial fluid composition. The length of *s* needed for RHS to be equal to LHS varies greatly among different flow paths, owing to the highly heterogeneous spatial distributions of flow velocity and the geometric surface.[25] In a chalk sample with a steady state flow field, a RHS variance of up to 6 orders of magnitude has been reported.[1] The dissolution front, as an isosurface, consists of all the points on each and every *s* that satisfies Equation 1, and is therefore morphologically complicated in 3D (Fig. 1a). A greater specific surface area (SSA) or longer residence time (s/v) yields shorter *s*.

Infiltration instability amplifies the transport heterogeneities in a porous medium by channelling the more reactive fluid toward the more permeable flow paths.[6] For example, regional permeability differences, often stemming from the local variations of material density, can be amplified by RII.[26] The solid-void interfaces in porous media are sharp spatial variations of material density.[27] They are also the surfaces for the fluid-solid interactions. Therefore, RII can increase the internal surface of a porous medium. In a simulation based on a 7×7×14 $\mu m^3$ volume of the microstructure of a real chalk sample (Fig. 1a), the geometric surface area (GSA) increased from 701.65 $\mu m^2$ to 1508.78 $\mu m^2$ in 450 s (Figs. 1b). This simulated percolation used an acidic solution (MilliQ pre-equilibrated with 1 bar $CO_2$ under 25 $^oC$, pH = 3.91) and a constant volumetric flowrate (5 $\mu m^3$/min). The GSA started decreasing after 450 s because of solid depletion. The same figure also shows that the non-monotonic microstructure evolution cannot be seen from the overall mass balancing. The solid residual, indicated by 1 – $\Phi$ (overall porosity), decreased monotonically while the centre of mass (CoM) in the axial direction of the simulation domain varies as function of time. First it moves towards the fluid outlet (positive displacement) because of preferential upstream dissolution, later it moves backwards because remaining solid near the fluid entrance was bypassed by the channelled fluid.

The flowing fluid approaches equilibrium faster with the increased reactive surface. As a result the dissolution front moves against the flow field toward the fluid entrance. This is demonstrated by the temporal pH mapping of the simulation domain (cross sections shown in Fig. 1c). The pH varies between 3.91 (initial) and 6.08 (equilibrium) and the dissolution front retraction took place between *t* = 0 (the instant when the microstructure is allowed to morph in the established flow and concentration fields) and 400 s. During this period, the region of the warmer colours (the more reactive subvolume of the domain) shrunk toward the left boundary because of the constantly increasing SSA on the RHS of Eq.1. Fig. 2a shows the dissolution rate[2] as a function of the axial position, averaged over the cross section. With the initial microstructure the dissolution rate decreased to $10^{-8}$ mol·m$^{-2}$·s$^{-1}$ within 3.5 μm. After 200 seconds, however, this distance moved, unexpectedly, opposite to the flow direction by 1.5 μm. At $t = 400$ s the position has retracted another 0.4 μm. This decrease in the retracting speed resulted from the depletion of solid near the fluid entrance (Fig. 2b). At $t = 600$ s the position advanced in the flow direction to 2.7 μm, indicating that the RII induced surface generation had been balanced out by the limited solid availability. This is in accordance with the observed change in the surface area distribution along the axial direction (Fig. 2c). At $t = 600$ s, the GSA was lower than in the previous timesteps near the inlet but at the same time still increasing between 8 to 12 μm. It is therefore expected that in a porous medium infinitely long in the flow direction, the reactive surface area at any given instant reflects a dynamic balancing between the RII-induced surface generation and the surface destruction caused by solid depletion. *i.e.*, the *in situ* surface area of a porous medium in a flow field is neither the inherent geometric surface nor necessarily less than the *ex situ* surface area. Hence, using *ex situ* geometric surface as the upper limit to estimate reactive surface can introduce great uncertainties in the analysis of water-rock interactions in a flow field.

**Retracting dissolution front recorded by X-ray imaging**

We used *in situ* X-ray microtomography (μCT) to record the dynamics of the microstructural evolution discussed above. A cylindrical chalk sample (ø×L = 0.9×2 mm) was used as a fast reacting model rock. The percolation fluid (MilliQ pre-equilibrated with 8 bar $CO_2$ at 50 °C) and flowrate (0.016 ml/min) was chosen such that the dissolution front is contained in the field of view (FOV) in μCT. The density of the porous material was reflected by the local X-ray linear absorption coefficient of the sample.[28] The sample's internal geometric surface was indicated by the spatial variations of the porosity.[27] Both quantities were measured in the intensity unit of the reconstructed greyscale images. The process was recorded continuously over a period of 88 hours (Fig. 3). Fig. 3a shows a monotonic decrease of the sample density, with a preferential dissolution region near the fluid entrance. Fig. 3b shows the evolution of the geometric surface along the flow direction. The values were the norms of the 3D intensity gradient vectors integrated over all the corresponding planes. A temporal increase of GSA was evident at $t = 28.5$ h (red). Fig. 3c shows the μCT images corresponding to the cross section 650 μm from the fluid inlet (indicated by the grey dashed line in Figs. 3a and 3b). A Sobel-Feldman operator was applied to the images to highlight the geometric surfaces.[29] A comparison between the images of 0 and 28.5 h indicates that the amplification of the pre-existing heterogeneities contributed considerably to the increase in surface area – a phenomenon characteristic to infiltration instability. Fig. 3d shows the evolution of the sample density (red circles) and the geometric surface (blue squares) averaged over the whole FOV (1×1×2 mm$^3$). Also shown is the displacement of the center of mass in the axial direction measured by X-ray absorption (columns). The dynamics of these quantities show a very good agreement with the model prediction (Fig. 1b). In the 4D dataset (the time series of a morphing 3D structure), the intensity in a timestep is subtracted from that of the following step, the difference reflects a spatial distribution of the solid removal rate. Fig. 3e shows this distribution in the axial direction for 5 pairs of consecutive timesteps. The smaller the value the more solid was removed. The shaded areas connected by the blue line show qualitatively the evolution of the most reactive portion of the sample. Retraction of the dissolution front toward the fluid inlet was observed in the first 3 time steps (pairs), in agreement with the increase of surface area shown in Figs. 3a and 3d. The later advancement of the shaded region resulted from the solid depletion that was directly observable from the image stacks (Supplementary files 1 & 2).

The unexpected retraction of dissolution front against the pressure gradient of the flow field strongly suggests the need to re-evaluate former determinations of dissolution kinetics. Accurate quantification of the reactive surface area is a prerequisite of studying many important subjects in water-rock interactions (*e.g.*, the Gibbs free energy dependence of reaction rate).[30] We have shown that the *in situ* reactive surface and the *ex situ* geometric surface can be dramatically different. Perhaps more surprising is that, given the same rock sample, the reactive surface in a flow field can be greater than its geometric surface without a flow field, even when the sample is chemically homogeneous and the precipitation of secondary phases is absent. This contradicts directly the long-held assumption that the geometric surface area constitutes the upper limit for the reactive surface area. This difference may also have contributed to the repeated reporting of discrepancies between laboratory and field site kinetic measurements.[31] In reality, many geologic settings are accompanied by divergent flows, *i.e.*, velocity decreases with the distance from the fluid source. If so, the natural porous media can often be considered infinitely long in the flow direction, and the balance between surface generation and removal determines the *in situ* reactive surface. To better quantify such complex dynamic processes, there is an urgent need to study the rate of RII-induced surface generation, the upper limit of the specific surface area reflected by the fractal dimension of geologic materials, as well as the rate of surface destruction dictated by solid availability.

**Methods**

**X-ray computerised tomography**

A model environment for numerical simulations was performed using the X-ray holotomography reconstructions[32] of drill cuttings from North Sea Hod chalk (Sample HC #16) with a diameter of ~500 μm. Imaging was performed at the ID22 beamline at the European Synchrotron Research Facility (ESRF) in Grenoble, France. We collected 1999 projections from a single dry scan at 29.5 keV with an exposure of 500 ms and 360º rotation. The reconstruction had a voxel size of 100 nm. For this study the reconstructions were intensity aligned, compensated for ring artifacts, denoised with iterative non-local means denoising[38] and sharpened by deconvolution. An estimate of voxel level porosity was acquired by interpolating from an intensity based Gaussian mixture model.[32]

Percolation experiments for the verification of the simulations were performed with outcrop Maastrichtian chalk collected near Aalborg, Denmark (Rørdal Quarry). The material contains primarily $CaCO_3$ in the form of calcite from fossilized coccoliths skeletal debris with ~ 4% silica content. The average porosity is high around 45%, permeability ranges between 3 and 5 mD and the BET surface was found to be 7.31 $m^2$/g.[33] We loaded our miniature version of a Hassler core holder[1] with an Aalborg chalk sample machined into a cylinder of 900 μm diameter and ~2 mm length. To guide the flow of solvent the chalk cylinder was wrapped with a heat shrinking tube (3M, Maplewood, MN, USA), placed between two stainless steel needles serving as fluid in- and outlet and sealed with epoxy resin prior to loading. The sample was confined by MilliQ water at 10 bar at the beginning of the experiment. The confining side was then sealed. A pump (260D Syringe Pump, Teledyne ISCO) was used to drive MilliQ water (pre-equilibrated with 8 bar $CO_2$ at 50 ºC) at 0.016 ml/min through the sample. The dissolution process was monitored by continuous imaging by X-ray microtomography (μCT) at the P05 beamline of PETRA III at the Deutsches Elektronen-Synchrotron (DESY). The beamline is operated by the Helmholtz-Zentrum Geesthacht.[34] Over the course of 88 hours 45 consecutive tomographic data sets were collected with a beam energy of 28 keV. Each set is consisting of 1200 projections exposed for 1050 ms over 180° rotation. The time resolution of was approximately 110 min. The effective voxel size in the reconstructions was 2.66 μm after downsampling the 3056×3056 pixel projection data by a factor of two in both directions. Prior to reconstruction the interface of the aluminum wall of the core holder to the confining water was tracked and matched with a reference sinogram to compensate for any drift and/or distortion caused by a changing dragging force of the applied tubing during rotation of the sample. Ring artifacts were suppressed by Fourier-wavelet based de-striping of the sinograms before reconstructing with the GridRec algorithm. After reconstruction we applied four iterations of iterative non-local means denoising to every timestep of the 4D dataset using a constant noise level estimate and aligned the dataset spatially with digital volume correlation using Pearson's correlation coefficient as the quality metric. Details of the signal processing can be found in Bruns et al.[32, 35]

**Greyscale model and simulation**

For numerical simulations we built a reactor network model that treats each voxel as a mixed flow reactor (MFR) and its connection to the 6-neighbourhood as an ensemble of plug flow reactors (PFR).[24] When a flow field is imposed the model allows relating the spatial distribution of voxel porosity with the reactor properties.[2] Within a 100 nm voxel we assumed a linear dependence of the permeability and tortuosity on the voxel level porosity, i.e. the pressure drop in the PFR, modeled by Darcy's law, uses a phenomenological coefficient in the form a parabolic function of local porosity given by the geometric mean of the connected reactors. The flow field was then solved by the stabilized biconjugate gradient method under the assumption of incompressible flow and continuity. The reactor model relates the reactant concentration with the mixing state in the reactor. Previous studies illustrated that the local conversion is determined by the voxel level Damköhler number.[36, 37, 38] A 50% subvolume of each voxel was treated as an MFR, avoiding the need to use effective diffusivities on the voxel level. An expression of surface area within each PFR is retrieved from the norm of the porosity gradient vector multiplied by the $2^{nd}$ power of the voxel size. The geometric surface area of each voxel is then evaluated by summation over all neighboring voxels. The nonlinear rate law of chalk dissolution[2] was solved based on the flow field iteratively until the Frobenius norm of the difference between two consecutive iterations was less than $10^{-6}$. Each iteration was initialized with the inlet concentration of the previous iteration. The simulation only considers transport of aqueous calcium, i.e. pH and saturation index in each reactor are calculated assuming calcite as the only available source of calcium in a closed compartment. Speciation calculations were performed separately using PhreeqC[39] and the voxel porosity was updated based on mass balancing after the chemical conversion in each reactor had been determined. The time step of the simulation was chosen adaptively, such that the overall porosity change would not exceed 0.01. For further details see Yang et al.[1]

**Author Contributions**

YY designed the research and built the model. YY, MR, SSH and JUH collected the *in situ* tomography data. SB processed the images. SLSS and HOS supervised the research. All authors discussed the results and commented on the manuscript.

**Acknowledgement**

We thank F. Saxild for help with the design and manufacturing of the percolation cell and H. Suhonen at the ID22 beamline at ESRF (European Synchrotron Research Facility, France) for technical support. We thank Imke Greving and Fabian Wilde for help with data collection at beamline P05 of DESY (Deutsches Elektronen-Synchrotron). Parts of this research were carried out at the light source PETRA III (Beamtime IDs I-20150242EC and I-20160208EC) at DESY, Germany, a member of the Helmholtz Association (HGF). Funding for this project was provided by the European Union's Horizon 2020 research and innovation programme under the Marie Sklodowska-Curie grant agreement No 653241, the Innovation Fund Denmark through the CINEMA project, as well as the Innovation Fund Denmark and Maersk Oil and Gas A/S through the P$^3$ project. Support for synchrotron beamtime was received from the Danish Agency for Science, Technology and Innovation via Danscatt.

# References

1. Yang Y, Hakim SS, Bruns S, Uesugi K, Stipp SLS, Sørensen HO. Wormholes grow along paths with minimal cumulative surface (under review). 2016.

2. Yang Y, Hakim SS, Bruns S, Rogowska M, Boehnert S, Hammel JU*, et al.* Direct observation of coupled geochemical and geomechanical impacts on chalk microstructural evolution under elevated CO2 pressure. Part I. (under review). 2016.

3. Yang Y, Bruns S, Stipp SLS, Sørensen HO. Dissolved CO2 stabilizes dissolution front and increases breakthrough porosity of natural porous materials (under review). 2016.

4. Yang Y, Bruns S, Stipp SLS, Sørensen HO. Reactive infiltration instability amplifies the difference between geometric and reactive surface areas in natural porous materials (under review). 2016.

5. Szymczak P, Ladd AJC. Reactive-infiltration instabilities in rocks. Fracture dissolution. *Journal of Fluid Mechanics* 2012, **702:** 239-264.

6. Chadam J, Hoff D, Merino E, Ortoleva P, Sen A. Reactive Infiltration Instabilities. *IMA Journal of Applied Mathematics* 1986, **36**(3)**:** 207-221.

7. Ortoleva P, Chadam J, Merino E, Sen A. Geochemical self-organization II: the reactive-infiltration instability. *Am J Sci* 1987, **287:** 1008-1040.

8. Noiriel C. Resolving Time-dependent Evolution of Pore-Scale Structure, Permeability and Reactivity using X-ray Microtomography. *Reviews in Mineralogy and Geochemistry* 2015, **80**(1)**:** 247-285.

9. Ellis BR, Peters CA. 3D Mapping of calcite and a demonstration of its relevance to permeability evolution in reactive fractures. *Advances in Water Resources* 2015.

10. Ellis BR, Fitts JP, Bromhal GS, McIntyre DL, Tappero R, Peters CA. Dissolution-driven permeability reduction of a fractured carbonate caprock. *Environmental engineering science* 2013, **30**(4)**:** 187-193.

11. Noiriel C, Luquot L, Madé B, Raimbault L, Gouze P, van der Lee J. Changes in reactive surface area during limestone dissolution: An experimental and modelling study. *Chemical Geology* 2009, **265**(1–2)**:** 160-170.

12. Noiriel C, Gouze P, Bernard D. Investigation of porosity and permeability effects from microstructure changes during limestone dissolution. *Geophysical Research Letters* 2004, **31**(24)**:** n/a-n/a.

13. Fredd CN, Fogler HS. Influence of transport and reaction on wormhole formation in porous media. *AIChE Journal* 1998, **44**(9)**:** 1933-1949.

14. Jasti J, Fogler HS. Application of neutron radiography to image flow phenomena in porous media. *AIChE journal* 1992, **38**(4)**:** 481-488.

15. Hoefner M, Fogler HS. Pore evolution and channel formation during flow and reaction in porous media. *AIChE Journal* 1988, **34**(1)**:** 45-54.

16. Hoefner ML, Fogler HS. Role of acid diffusion in matrix acidizing of carbonates. *Journal of petroleum technology* 1987, **39**(02)**:** 203-208.

17. Rege SD, Fogler HS. Competition among flow, dissolution, and precipitation in porous media. *AIChE Journal* 1989, **35**(7)**:** 1177-1185.

18. Zhao C, Hobbs B, Ord A. Computational simulation of chemical dissolution-front instability in fluid-saturated porous media under non-isothermal conditions. *International Journal for Numerical Methods in Engineering* 2015, **102**(2)**:** 135-156.

19. Zhao C, Hobbs B, Ord A. Theoretical analyses of chemical dissolution-front instability in fluid-saturated porous media under non-isothermal conditions. *International Journal for Numerical and Analytical Methods in Geomechanics* 2015, **39**(8)**:** 799-820.

20. Zhao C, Hobbs BE, Ord A. Effects of medium and pore-fluid compressibility on chemical-dissolution front instability in fluid-saturated porous media. *International Journal for Numerical and Analytical Methods in Geomechanics* 2012, **36**(8)**:** 1077-1100.

21. Szymczak P, Ladd AJ. Reactive-infiltration instabilities in rocks. Part 2. Dissolution of a porous matrix. *Journal of Fluid Mechanics* 2014, **738:** 591-630.

22. Szymczak P, Ladd AJ. Instabilities in the dissolution of a porous matrix. *Arxiv preprint arXiv:11033816* 2011.

23. Szymczak P, Ladd A. Wormhole formation in dissolving fractures. *Journal of Geophysical Research: Solid Earth (1978–2012)* 2009, **114**(B6).

24. Fogler HS. *Elements of Chemical Reaction Engineering*. Pearson Education, 2016.

25. Kalia N, Balakotaiah V. Effect of medium heterogeneities on reactive dissolution of carbonates. *Chemical Engineering Science* 2009, **64**(2)**:** 376-390.

26. Hinch E, Bhatt B. Stability of an acid front moving through porous rock. *Journal of Fluid Mechanics* 1990, **212:** 279-288.

27. Yeong CLY, Torquato S. Reconstructing random media. *Physical Review E* 1998, **57**(1)**:** 495-506.

28. Cnudde V, Boone M. High-resolution X-ray computed tomography in geosciences: A review of the current technology and applications. *Earth-Science Reviews* 2013, **123:** 1-17.

29. Sobel I, Duda R, Hart P, Wiley J. Sobel-Feldman Operator.

30. Brantley SL, Kubicki JD, White AF. *Kinetics of water-rock interaction*, vol. 168. Springer, 2008.

31. Maher K, Steefel CI, DePaolo DJ, Viani BE. The mineral dissolution rate conundrum: Insights from reactive transport modeling of U isotopes and pore fluid chemistry in marine sediments. *Geochimica et Cosmochimica Acta* 2006, **70**(2)**:** 337-363.

32. Bruns S, Sørensen HO, Stipp SLS. Rock Properties of Compacted North Sea Chalks characterized by Greyscale Analysis. *Water Resources Research (under review)* 2016.

33. Okhrimenko DV, Dalby KN, Skovbjerg LL, Bovet N, Christensen JH, Stipp SLS. The surface reactivity of chalk (biogenic calcite) with hydrophilic and hydrophobic functional groups. *Geochimica et Cosmochimica Acta* 2014, **128:** 212-224.

34. Wilde F, Ogurreck M, Greving I, Hammel JU, Beckmann F, Hipp A*, et al.* Micro-CT at the imaging beamline P05 at PETRA III.  AIP Conference Proceedings; 2016: AIP Publishing; 2016. p. 030035.

35. Bruns S, Stipp SLS, Sørensen HO. Looking for the Signal: A Guide to Iterative Noise and Artefact Removal in X-ray Tomography Reconstructions of Reservoir Rocks. *Water Resources Research (under review)* 2016.

36. Levenspiel O. *Chemical reaction engineering*. Wiley & Sons, Inc., 1999.

37. Zwietering TN. The degree of mixing in continuous flow systems. *Chemical Engineering Science* 1959, **11**(1)**:** 1-15.

38. Danckwerts P. The effect of incomplete mixing on homogeneous reactions. *Chemical Engineering Science* 1958, **8**(1)**:** 93-102.

39. Parkhurst DL, Appelo C. User's guide to PHREEQC (Version 2): A computer program for speciation, batch-reaction, one-dimensional transport, and inverse geochemical calculations. 1999.

**Figure legends**

**Figure 1**. Reactive transport simulation based on a greyscale nanoCT dataset of natural chalk. (a) A perspective view of the initial microstructure (7×7×14 µm$^3$). The instantaneous reaction front is imposed on the structure as a yellow isosurface. Fluid flows from left to right. Above the structure shows a cross section of the distribution of calcite saturation index in the initial flow field. The dissolution front separates the yellow region (SI$_{calcite}$ = 0) from the rest of the domain. At the bottom are images showing the instantaneous geometric and reactive surfaces of the same cross section. Only the geometric surface contained by the reaction front is reactive. (b) The evolution of geometric surface (blue squares) and solid content (1 - $\Phi$) of the simulation domain (red circles). Also shown is the change in the centre of mass (columns), which reflects the distribution of solid along the flow axis. (c) The evolution of the pH distribution. The same axial cross section in (a) is used. The initial pH of the solution is 3.91. The migration of the dissolution front in the opposite direction of the pressure gradient was observed from $t$ = 0 to 400 s.

**Figure 2**. Evolution of the axial heterogeneities in the numerical simulation. The origin of the x-axis is the fluid inlet. (a) Distribution of chalk dissolution rate along the flow direction. The rate is computed based on a related work.$^2$ (b) Distribution of solid material along the axial direction, whereas $\Phi$ represents the porosity averaged over the plane perpendicular to the flow. (c) Distribution of geometric surface area (GSA) along the axial direction.

**Figure 3**. Evolution of chalk microstructure recorded by *in situ* X-ray tomography. (a) Axial distribution of X-ray absorption. The absorption reflects the average density of the sample in the radial direction. The strong absorption near the fluid inlet was caused by a stainless steel (SS) tube. (b) Axial distribution of geometric surface area, calculated as the norms of the 3D intensity

gradient vectors integrated over the radial plane. (c) Cross sections of the evolving microstructure 650 μm away from the fluid inlet (the grey dash line in a & b). The edges are highlighted using the Sobel-Feldman operator. The pre-existing structural heterogeneities were greatly enhanced in 28.5 hrs. (d) Evolution of X-ray absorption (red circles), geometric surface (blue squares) and the displacement of centre of mass (columns, measured by X-ray absorption) of the complete field of view (FOV). There was an interruption of fluid pumping during the *in situ* measurement at around $t = 60$ hour. (e) Decrease of X-ray absorption between timesteps as a function of axial position. The shaded areas show the regions where the most solid was removed by the fluid during the consecutive scans.

**Figure 1**

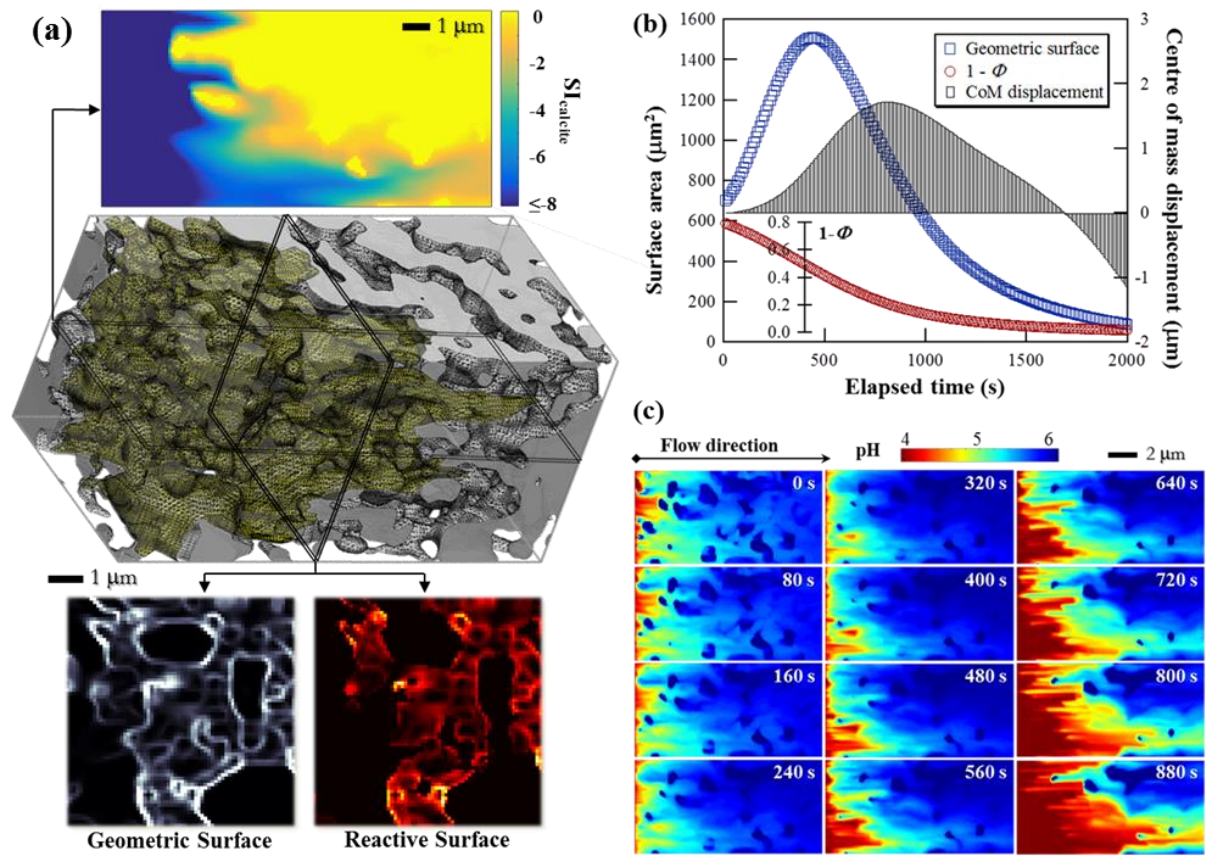

**Figure 2**

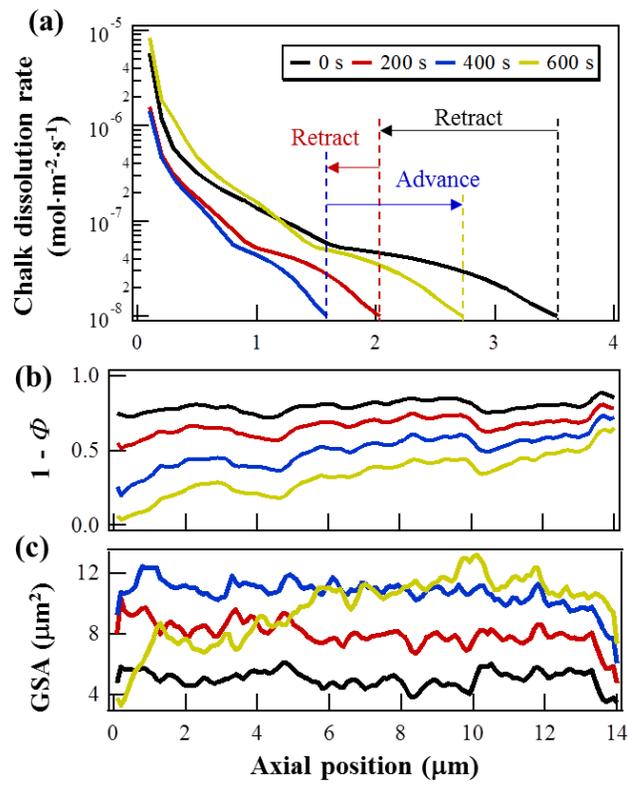

**Figure 3**

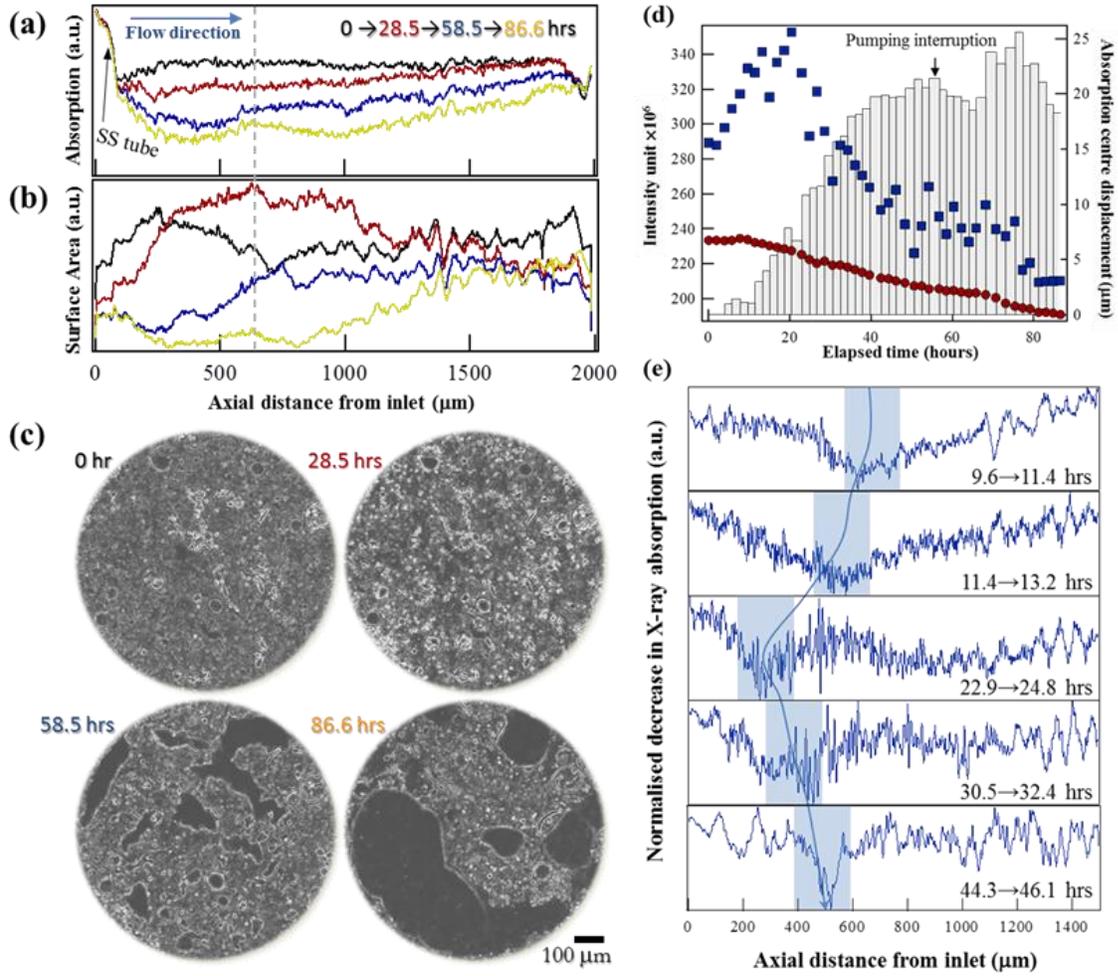